\documentclass[sigconf]{acmart}
\AtBeginDocument{%
  }

\setcopyright{acmlicensed}
\copyrightyear{2026}
\acmYear{2026}
\acmDOI{XXXXXXX.XXXXXXX}
\acmConference[CHI '26]{2026 CHI Conference on Human Factors in Computing Systems}{April 13--17,
  2026}{Woodstock, NY}
\acmISBN{978-1-4503-XXXX-X/2018/06}

\acmConference[PoliSim@CHI 2026]{PoliSim@CHI 2026: LLM Agent Simulation for Policy}{April 16, 2026}{Barcelona, Spain}
\renewcommand{\footnotetextcopyrightpermission}[1]{%
  \footnotetext{%
    This paper was prepared for \href{https://polisim.net/}{PoliSim@CHI 2026: LLM Agent Simulation for Policy}, a workshop at \href{https://chi2026.acm.org/}{CHI 2026 (CHI Conference on Human Factors in Computing Systems)}, April 16, 2026, Barcelona, Spain.
  }%
}

\begin{document}

\title{SLALOM: Simulation Lifecycle Analysis via Longitudinal Observation Metrics for Social Simulation}


\author{Juhoon Lee}
\affiliation{%
  \institution{School of Computing, KAIST}
  \city{Daejeon}
  \country{Republic of Korea}}
\email{juhoonlee@kaist.ac.kr}

\author{Joseph Seering}
\affiliation{%
  \institution{School of Computing, KAIST}
  \city{Daejeon}
  \country{Republic of Korea}}
\email{seering@kaist.ac.kr}

\renewcommand{\shortauthors}{Lee et al.}

\begin{abstract}
Large Language Model (LLM) agents offer a potentially-transformative path forward for generative social science but face a critical crisis of validity. Current simulation evaluation methodologies suffer from the ``stopped clock'' problem: they confirm that a simulation reached the correct final outcome while ignoring whether the trajectory leading to it was sociologically plausible. Because the internal reasoning of LLMs is opaque, verifying the ``black box'' of social mechanisms remains a persistent challenge. In this paper, we introduce SLALOM (Simulation Lifecycle Analysis via Longitudinal Observation Metrics), a framework that shifts validation from outcome verification to process fidelity. Drawing on Pattern-Oriented Modeling (POM), SLALOM treats social phenomena as multivariate time series that must traverse specific \textit{SLALOM gates}, or intermediate waypoint constraints representing distinct phases. By utilizing Dynamic Time Warping (DTW) to align simulated trajectories with empirical ground truth, SLALOM offers a quantitative metric to assess structural realism, helping to differentiate plausible social dynamics from stochastic noise and contributing to more robust policy simulation standards.
\end{abstract}


\keywords{Simulation Evaluation, Pattern-Oriented Modeling, Agent-based Modeling}

\received{12 February 2026}
\received[revised]{12 February 2026}
\received[accepted]{19 March 2026}

\maketitle

\section{Introduction}
Simulations are useful for understanding the real world. It is no surprise that generative social science has long sought to grow artificial societies from the bottom up to explain macroscopic social phenomena~\cite{epstein2012generative}. The recent advent of Large Language Model (LLM) agents offers a potentially-transformative leap for this field~\cite{park2023generative, park2022social}. Unlike traditional Agent-Based Models (ABMs) constrained by rigid heuristic rules, LLM agents possess semantic richness and are capable of interacting and evolving in ways that mimic human social complexity.

However, this expressivity introduces a profound validation challenge. Critics argue that LLM-based simulations are ``black boxes'' built upon ``black boxes''~\cite{larooij2025validation, taillandier2025integrating}. Because the internal reasoning of an LLM is opaque and stochastic, we cannot easily verify that a simulation is operating on sound sociological principles. If a simulated population mimics the impact of a policy, was it because it captured the realistic dynamics of the situation, or simply because of stochastic hallucinations?

To address this, researchers navigate a fragmented methodological landscape. A recent review of Generative AI simulation practices by Larooij and Törnberg \cite{larooij2025validation} catalogs five distinct validation strategies, ranging from comparisons against human-generated data and internal consistency checks to expert judgment. Despite this methodological breadth, they observe that in practice, the field often defaults to subjective face validity or the reproduction of static stylized facts. While these approaches confirm that agents sound plausible or match broad statistical generalizations, they generally fail to verify that the underlying social processes are robust over time.

Turning to the broader Agent-Based Modeling (ABM) community, Collins et al. \cite{collins2024methods} provide a comprehensive methodological taxonomy, detailing nine distinct validation categories that span from qualitative role-playing to data analytics and empirical validation. However, a gap remains when applying these established methods to generative agents. Traditional empirical validation often defaults to point-matching --- comparing the simulation’s final aggregate output to real-world data. This creates the ``stopped clock'' problem: a simulation might reach the correct final state via a completely wrong trajectory. For example, a simulated public may behave properly, matching the expected outcome, but due to apathy rather than the intended intervention. For social science, verifying the outcome is insufficient; we must also validate the process.

In this paper, we argue that validating the ``black box'' of LLM agents requires a shift toward temporal pattern matching. We draw inspiration from Pattern-Oriented Modeling (POM) in theoretical ecology~\cite{grimm2005pattern, gallagher2021theory}. Ecologists validate complex biological systems not by dissecting their atomic mechanics, but by requiring the model to simultaneously match multiple structural patterns across different temporal scales.

We introduce SLALOM (Simulation Lifecycle Analysis via Longitudinal
Observation Metrics), a framework for the phasic evaluation of simulation validity. It functions by transforming the unstructured textual output of agent interactions into multivariate time series, analyzing dynamic variables such as sentiment, volatility, and diversity. We define \textit{SLALOM gates} as intermediate waypoint constraints representing the distinct phases of a social phenomenon. Crucially, SLALOM does not rely on a single metric; instead, it performs Dynamic Time Warping (DTW)~\cite{berndt1994dynamic} across these multiple dimensions simultaneously. By aggregating the alignment scores of these diverse variables against empirical ground truth, SLALOM validates the fidelity of social change over time, capturing the full complexity of the social dynamic rather than overfitting to a single output curve.

\section{Related Works}

\subsection{Agentic Simulations in Social Science} 

Agent-Based Models (ABMs) are the computational backbone of generative social science~\cite{epstein2012generative}, allowing researchers to grow macro-phenomena from micro-rules, from Schelling’s foundational segregation models~\cite{schelling1971dynamic} to modern epidemiological simulations~\cite{kerr2021covasim, will2020combining}. However, traditional ABMs were constrained by the heuristic bottleneck: agents followed rigid rules, limiting their ability to model complex human behaviors.

The integration of Large Language Models (LLMs) has fundamentally shifted this paradigm. Works such as Generative Agents~\cite{park2023generative} and Social Simulacra~\cite{park2022social} demonstrated that agents could possess semantic richness, forming memories, relationships, and daily routines that mimic human social complexity with believable fidelity. This shift from numerical parameters to natural language interaction offers a new frontier for policy modeling. However, recent critiques highlight a critical validity gap~\cite{larooij2025validation}. While these agents exhibit high face validity and sound human, they often fail to replicate complex and more nuanced structural behaviors, such as irrational groupthink, sub-optimal coordination, or long-term relationship decay, leading to what some have termed ``stochastic parrots'' in social clothing.

\subsection{The Validation Crisis} 

Validating these black-box models remains an important challenge. As reviewed by Larooij and Törnberg, current practices rely heavily on subjective measures like human expert judgment or Turing-style evaluations~\cite{larooij2025validation}, which are unscalable and prone to bias. Conversely, rigorous benchmarks like AgentBench~\cite{liu2023agentbench} or WebArena~\cite{zhou2023webarena} focus on binary task completion (e.g., ``Did the agent book the flight?'') rather than social processes.

While the broader ABM community offers rigorous validation frameworks, such as the nine methods outlined by Collins et al.~\cite{collins2024methods}, these often default to point-matching final aggregate statistics. This ignores the \textit{trajectory} of the phenomenon; a simulation might reach the correct final state via a completely unrealistic path, a failure mode that standard benchmarks miss.

\subsection{Pattern-Oriented Modeling (POM)} 

To address this, we look to theoretical ecology. Grimm et al.~\cite{grimm2005pattern} proposed Pattern-Oriented Modeling (POM) to filter out structurally invalid models. The core philosophy is that matching a single output variable is insufficient; a valid model must simultaneously reproduce multiple patterns observed at different scales.

This approach has successfully validated complex ecological systems, such as fish schooling~\cite{huth1992simulation}, where simple distance metrics failed to distinguish between competing theories. Recently, Wang et al.~\cite{wang2024known} extended POM to handle ``unknown unknowns'' in modeling, and Epstein~\cite{epstein2023inverse} proposed Inverse Generative Social Science (IGSS) to automatically discover rules that fit multi-dimensional patterns. We adapt a similar framework to the social ecosystem of LLM agents, treating the SLALOM gates as the essential patterns that define structural validity.
\section{Assumptions}
To establish the SLALOM framework, we posit three core assumptions.

\subsection{Assumption 1: Phasic Archetypes}
We assume that complex social phenomena (e.g., polarization, panic diffusion, norm formation) are not random walks but follow archetypal temporal structures. For example, as defined in classical crisis management theory, a crisis typically follows a recurring lifecycle: prodromal, acute, chronic, and resolution~\cite{fink1986crisis}. Similarly, anthropological accounts of ``social dramas'' describe a universal four-phase progression from breach to reintegration~\cite{turner1980social}, emphasizing semiotic approaches~\cite{wirtz2023social}. We assume these structural phases exist in social phenomena, even if specific content varies.

\subsection{Assumption 2: Observable Temporal Social Signals} 
We assume that the internal state of a black-box agent society can be inferred from its text trace. Following the principles of computational social science~\cite{lazer2009life}, we treat interaction logs as valid proxies for social behavior. Research in linguistic psychology confirms that aggregate word usage patterns reliably reflect underlying cognitive and emotional states~\cite{pennebaker2003psychological}, allowing us to convert text into time-series data. We posit that latent variables such as anxiety, cohesion, or informational entropy can be extracted from interaction logs via natural language processing (NLP) techniques, such as sentiment analysis or embedding distance.

\subsection{Assumption 3: Trajectory Validity}
We assume that generative sufficiency~\cite{epstein2012generative} for policy analysis does not require exact replication of all aspects of a social phenomenon. Instead, a simulation is valid if its trajectory passes through the same ``validity regions'' (\textit{SLALOM gates}) as the empirical data. We adopt the POM perspective~\cite{grimm2005pattern}, which argues that a model is structurally realistic if it simultaneously reproduces multiple observed patterns at different scales. If a simulation navigates the same sequence of structural states as the real world, it captures the essential mechanisms of the phenomenon.
\section{SLALOM Framework}
We propose SLALOM, a framework for evaluating agentic simulations via multivariate trajectory matching with waypoint constraints.

\subsection{SLALOM Gates}
The parameter space of an LLM-based society is effectively infinite and dominated by stochastic noise. Exploring this full space is computationally intractable. SLALOM addresses this by treating validity gates as a structural pruning mechanism.

A SLALOM gate is defined as a tuple $\{t_{window}, V_{min}, V_{max}, M_k\}$. Crucially, these act as binary filters: if a trajectory misses a gate, it is pruned from the analysis. This restricts the evaluation to the ``sociological near-neighbors'' of the empirical ground truth. Rather than accepting any path that arrives at the destination, SLALOM enforces that the simulation must reside within the valid bounded region of the parameter space, effectively filtering out sociologically incoherent variations before they propagate to the final state.

\subsection{Evaluation Metric: Aggregate Dynamic Time Warping}

Social time is elastic; a simulated discussion might resolve in 50 turns where humans take 100. Furthermore, different social variables may unfold at different speeds. Standard Euclidean distance would penalize these natural temporal shifts heavily. We employ DTW to measure the similarity between the Simulation Trajectory ($S$) and the Target Trajectory ($T$). DTW finds an optimal alignment between two time-dependent sequences by warping the time axis to minimize the distance. Formally, for two sequences $S$ and $T$, the DTW distance is defined as:

\begin{equation}
DTW(S, T) = \min_{W} \left[ \sum_{k=1}^{K} \delta(w_k) \right]
\end{equation}

where $\delta(w_k)$ is the distance between the aligned data points at step $k$ of the path, and the minimization is performed over all possible monotonic warping paths $W$. A low DTW score indicates that the simulation ``hit the SLALOM gates'' in the correct order and relative duration, despite temporal shifts. To evaluate the overall validity of the simulation, SLALOM aggregates the normalized DTW scores across all $K$ variable dimensions: 
$$Score_{total} = \sum_{k=1}^{K} w_k \cdot DTW(S_k, T_k)$$ 
A low aggregate DTW score indicates that the simulation proceeded in the correct order and relative duration across multiple dimensions, validating the causal structure of the events rather than just the final output.
\section{Case Study: Small Group Dynamics}
To validate the SLALOM framework, we applied it to small group product design. We mathematically constructed validity gates from the AMI Meeting Corpus~\cite{kraaij2005ami}, establishing a ground truth for how valid human teams navigate the Tuckman Developmental Sequence (Forming, Storming, Norming, Performing)~\cite{tuckman1965developmental}.

\subsection{Ground Truth Processing}
We utilized the Scenario subset of the AMI corpus (15 groups, 4 members each). To create a cohesive longitudinal baseline, we concatenated the four sequential design meeting phases (A--D) for each group into a single continuous timeline, cutting the beginning and ending 5\% of B to D to minimize noise (e.g., greetings) introduced when a meeting starts and ends for subsequent meetings. We then normalized all timestamps to a percentage scale $t \in [0, 100]$ and discretized the timeline into $B=100$ bins. For each bin, we aggregated the interaction logs to calculate the mean ($\mu_{GT}$) and standard deviation ($\sigma_{GT}$) for three factors: (1) \textbf{Hierarchy} ($G_t$, Gini of word counts~\cite{dorfman1979formula, lewis2023instructor}) to measure speaker dominance; (2) \textbf{Divergence} ($D_t$, SBERT Divergence~\cite{reimers2019sentence}) to measure conceptual diversity; and (3) \textbf{Cohesion} ($LSM$, Language Style Matching~\cite{gonzales2010language}) to proxy implicit cohesion.

\subsection{Defining the Statistical Validity Gates}
We define a SLALOM Gate as a probabilistic region $Gate_t = [\mu_{GT}(t) \pm 2\sigma_{GT}(t)]$, capturing the 95\% confidence interval of valid behavior. 

We mapped these regions to the Tuckman Sequence:

\begin{enumerate}
    \item \textbf{Forming ($t \approx 0.25$):} Valid groups exhibit speaker hierarchy ($G_t \approx 0.48$). Divergence is moderate ($D_t \approx 0.3$).
    \item \textbf{Storming ($t \approx 0.45$):} Participation becomes more distributed as the team engages with each other, causing the Gini coefficient to drop ($G_t \approx 0.37$). Linguistic coordination begins to spike ($LSM \uparrow 0.4$).
    \item \textbf{Norming ($t \approx 0.7$):} Language Style Matching continues increasing ($LSM \approx 0.5$), signaling the successful formation of a shared group identity. Participation remains distributed ($G_t \approx 0.32$).
    \item \textbf{Performing ($t \approx 0.98$):} Cohesion relaxes slightly but remains high ($LSM \approx 0.42$), while the hierarchy stabilizes ($G_t \approx 0.35$) as the team executes the plan.
\end{enumerate}

\begin{figure*}[t]
    \centering
    \includegraphics[width=\textwidth]{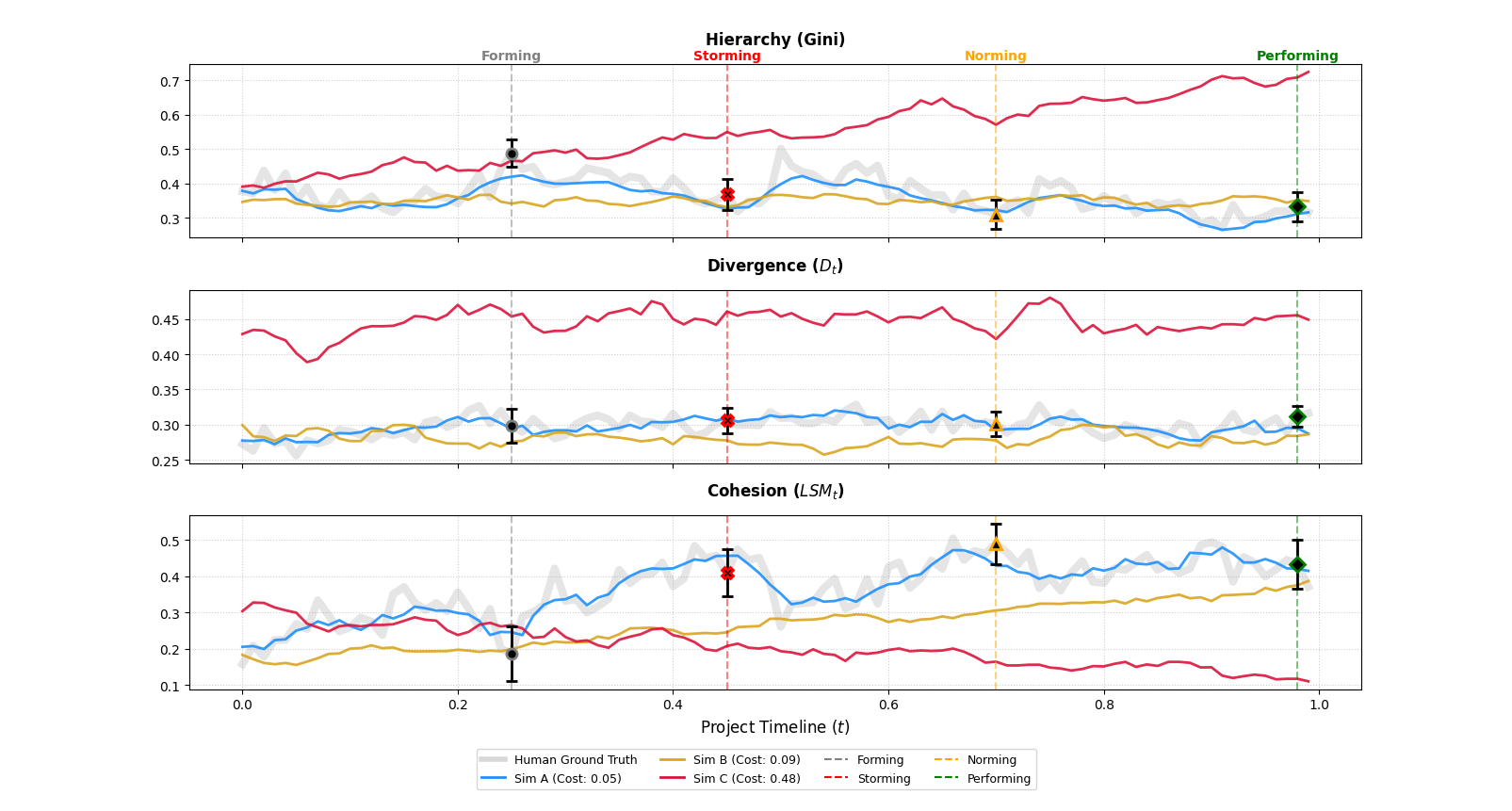}
    
    \caption{\textbf{SLALOM Validation Results.} 
    The gray region represents the AMI Ground Truth ($\mu_{GT} \pm 2\sigma_{GT}$).}
    \label{fig:slalom_results}
\end{figure*}

\subsection{Diagnostic Application: Synthetic Traces}

To demonstrate SLALOM’s discriminative power, we applied these gates to three hypothetical simulation trajectories (Sim A, B, C) and calculated their alignment with the AMI ground truth using DTW (Table 1). Sim C exhibits clear failure: while it successfully generated divergence, it spiraled into runaway dominance ($G_t \rightarrow 0.7$) and crashing cohesion ($LSM \downarrow 0.1$), resulting in the highest total validity cost (0.480). This flags a catastrophic failure in social mechanisms of the simulation. Conversely, Sim B suffered from stagnation, maintaining a relatively flat hierarchy and divergence profile that failed to capture the necessary Storming volatility, reflected in a moderate validity cost (0.096). Only Sim A achieved a low total cost (0.049), demonstrating the necessary phase transitions---establishing hierarchy to manage conflict, then building cohesion to execute the vision---required for structural realism.

\begin{table}[h]
    \centering
    \caption{\textbf{DTW Validity Costs.} Lower scores indicate better structural alignment with human teams.}
    \label{tab:dtw_scores}
    
    \resizebox{\columnwidth}{!}{%
    \begin{tabular}{lcccc}
        \toprule
        \textbf{Sim} & \textbf{Hierarchy} & \textbf{Divergence} & \textbf{Cohesion} & \textbf{Total} \\
        \midrule
        A & \textbf{0.018} & \textbf{0.006} & \textbf{0.024} & \textbf{0.049} \\
        B & 0.031 & 0.013 & 0.052 & 0.096 \\
        C & 0.159 & 0.151 & 0.171 & 0.480 \\
        \bottomrule
    \end{tabular}%
    }
    
    \vspace{1mm}
\end{table}
\section{Discussion}

\subsection{Macro-level Accuracy}
SLALOM shifts policy simulation from outcome optimization to process safety, measuring macro-level accuracy. Traditional metrics might deem a simulation successful if it achieves a target (e.g., Reduce toxicity by 20\%), regardless of whether this result was reached via healthy dialogue or censorship. By enforcing intermediate waypoint constraints, SLALOM acts as a forensic tool to distinguish these mechanisms, allowing policymakers to audit unintended consequences before deployment.

This approach addresses the ``black box'' opacity of LLMs by arguing that strict mechanistic interpretability is unnecessary if the simulation demonstrates \textit{structural realism}~\cite{grimm2005pattern} to achieve generative sufficiency~\cite{epstein2012generative}. If agents spontaneously generate the correct sequence of macro-level phases, they are likely capturing essential sociolinguistic dynamics, effectively bounding the model's hallucination space to sociologically plausible trajectories. For policy design, the mechanism is the metric. If a simulation achieves a 20\% reduction in toxicity by ``silencing minority voices'' rather than ``fostering dialogue'', it is dangerous for policymakers to rely on it, even if the final number is correct. SLALOM acts as a forensic tool for these mechanisms.

\subsection{Limitations}
We note several critical limitations with our framework. First, SLALOM’s effectiveness is contingent on the resolution of the target data; without high-frequency longitudinal ground truth, which is often scarce in social science, the definition of precise validation gates becomes arbitrary. Second, the use of DTW assumes a monotonic temporal progression. Consequently, this metric cannot accurately evaluate simulations that exhibit radical branching, looping topologies, or non-linear social time, where the structural sequence of events diverges fundamentally from the ground truth.
\section{Conclusion}

Validation remains the primary bottleneck preventing Generative AI from revolutionizing computational social science. As long as simulations are evaluated solely on their static outputs, they remain susceptible to the ``stopped clock'' fallacy --- arriving at the correct social outcome via a hallucinated or sociologically invalid trajectory. In this paper, we introduced SLALOM, a framework that shifts the paradigm from outcome verification to process fidelity. By translating the principles of Pattern-Oriented Modeling (POM) into the domain of Large Language Models, SLALOM provides a rigorous method to audit the ``black box'' of agentic society. We argue that by enforcing multivariate waypoint constraints, we can mathematically distinguish between stochastic parroting and genuine structural realism. By validating the longitudinal geometry of social change, SLALOM aims to convert generative agents from fascinating toys into reliable, auditable instruments for policy research.

\bibliographystyle{ACM-Reference-Format}
\bibliography{sample-base}

\appendix

\end{document}